\documentclass[twocolumn,showpacs,prl]{revtex4}

\usepackage{graphicx}%
\usepackage{dcolumn}
\usepackage{amsmath}
\usepackage{latexsym}

\begin {document}

\title{Scale-free Network on Euclidean Space Optimized by Rewiring of Links}

\author
{
S. S. Manna$^{1,2}$, A. Kabak\c c\i o\u glu$^1$
}
\affiliation{
$^1$INFM - Dipartimento di Fisica, Universit\`a di Padova, I-35131 Padova, Italy \\
$^2$Satyendra Nath Bose National Centre for Basic Sciences Block-JD, Sector-III, Salt Lake, Kolkata-700098, India
}

\begin{abstract}    
  A Barab\'asi-Albert scale-free network is constructed whose 
  nodes are the Poisson distributed random points within a unit square and 
  links are the straight line connections among the nodes. 
  The cost function, which is the total wiring length associated with a
  such a network defined on a two dimensional plane is optimized.
  The optimization process consists of random selection of a pair of links and rewiring them
  to reduce the total length of the pair but with the constraint that the 
  degree as well as the out-degree and in-degree of each node are precisely 
  maintained. The resulting optimized network has a small diameter as well
  as high clustering and the link length distribution has a stretched 
  exponential tail.

\end{abstract}
\pacs {02.50.-r 
	89.20.-a 
	89.75.-k 
	89.75.Hc 
}
\maketitle

    The Internet is a very complex network connecting the large number of 
  computers around the world \cite {Faloutsos,web}. The nodes of this network may be interpreted
  as the routers and links as the cables connecting computers. The network
  can also be described in the inter-domain level where each domain is 
  represented by a single node and each link is an inter-domain connection.
  Such a network is well described by a graph consisting of a set of vertices
  and another set of edges among the vertices. Without assigning any weight 
  with the links between the nodes only the topological structure of the
  Internet is meaningful. Study of Internet's topological structure may be
  important for designing efficient routing protocols and modeling Internet traffic.

    Internet is one of the large class of real-world networks that exhibit
  small-world and/or scale-free properties, e.g.,  social networks \cite {Newman}, 
  biological networks \cite {Jeong,Sole}, electronic communication networks \cite {Faloutsos,web} etc.
  Quantities that characterize a network of $N$ nodes are the diameter ${\cal D}(N)$
  which measures the topological extension of the network, the clustering co-efficient
  ${\cal C}$ measures the local correlations among the links of the network and the nodal degree 
  distribution ${\cal P}(k)$. In a small-world network (SWN) \cite{WS}, the diameter 
  ${\cal D}(N)$ of the network scale logarithmically with $N$ 
  where as for a scale-free network (SFN) the degree distribution has a power 
  law tail: ${\cal P}(k) \propto k^{-\gamma}$. Barab\'asi and Albert (BA) showed that a
  growing network with preferential link attachment probability is a SFN with
  $\gamma=3$ \cite{BA}.

    Waxman first studied probabilistic graph models of Internet where links have
  weights which are their physical lengths \cite {Waxman}. The link length distribution in
  such networks decays exponentially: $D(\ell) \sim \exp(-\ell/\ell_o)$.
  Faloutsos et. al. observed that the out-degree distribution of Internet follows
  a power law tail \cite {Faloutsos}. Yook et. al. observed that the distribution
  of routers of North America is a fractal set and the link length distribution is
  inversely proportional to the link lengths \cite {Yook}. It is 
  suggested that in the growing Internet when a new node is becoming member
  of the network two competing factors control the decision to which node of
  the already grown Internet the new node will be connected. The factors are the
  degree $k_i$ of the existing node $i$ and in general the $\alpha$-th power of the length $\ell$ of the link 
  connecting the new node and the node $i$. The preferential attachment
  probability for the $i$-th node is therefore: $\pi_i \propto k_i \ell^{\alpha}$.

\begin{figure}[top]
\begin{center}
\includegraphics[width=6.0cm]{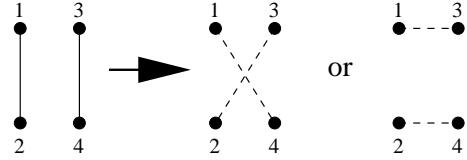}
\end{center}
\caption{
Two possible rewirings of a pair of links to reduce the total length of the two links.
}
\label{Fig1}
\end{figure}

    Recently it has been argued that such a network is scale-free for all values
  of $\alpha > \alpha_c = 1 - d$ in $d$ dimension and the link length distribution 
  generally follows a power law $D(\ell) \sim \ell^{\delta}$ where $\delta(\alpha) 
  = \alpha + d - 1$ \cite {MS,MS1}. For  $\alpha < \alpha_c$ the degree distribution 
  decays stretched exponentially but $D(\ell)$ still maintains a power law where 
  $\delta$ saturates at $-d-1$. The limit of $\alpha \to -\infty$ is interesting
  where each node connects only to its nearest earlier node. In a regular network
  in the form of a linear chain similar studies have been done \cite {Xulvi}.
  An interplay between the preferential attachment and the link length selection 
  within an interaction range for the Euclidean networks
  is studied in \cite {Barthelemy}.

    In this paper we associate a cost function associated with such networks.
  Each link of the network has the cost equal to its Euclidean length $\ell$ and
  therefore the cost function of the whole network is the total length of all the links
  of the network. The question we ask is, how can one construct a small-world scale-free 
  network with minimal cost? To study this we start generating a $N$ 
  node BA SFN on a two-dimensional plane. Links are then interchanged to reduce 
  the cost function keeping the topology i.e. the degree value of each node intact.
  The optimization of the wiring length of networks on lattices has been studied 
  in \cite {Rozenfeld}.

    We start with constructing a Barab\'asi-Albert SFN embedded in the Euclidean space
  as follows. Let $(x_1, x_2, ... , x_N)$ and $(y_1, y_2, ... , y_N)$ be the independent 
  identically and uniformly distributed random variables on the interval [0,1]. To 
  construct one random configuration of the network let a specific set of values of the 
  $N$ pair variables $\{(x_1,y_1), (x_2,y_2), ... , (x_N,y_N)\}$ be the co-ordinates 
  of the set of $N$ points on the unit square representing the set of nodes of the network 
  with serial numbers $i=1$ to $N$ assigned to them. We use first $m+1$ points 
  with serial numbers $i=1$ to $m+1$ to construct a $m+1$-clique by connecting each point 
  with rest of the $m$ points. Then following the serial numbers new points are
  added to the network one after another and each node is connected to randomly 
  selected $m$ distinct previous nodes.
  The probability to link the new node with serial number $j$ to a previous 
  node $i$ is linearly proportional to its degree, $k_i$. The network 
  thus constructed up to $N$ nodes is exactly the BA network \cite {BA}.
  At the same time it is a small-world network, i.e. the diameter ${\cal D}(N)$ 
  of the network measured by the maximal distance between an arbitrary pair of 
  nodes grow as  $\log(N)$. In this paper we restrict ourselves to $m=2$.

\begin{figure}[top]
\begin{center}
\includegraphics[width=8.0cm]{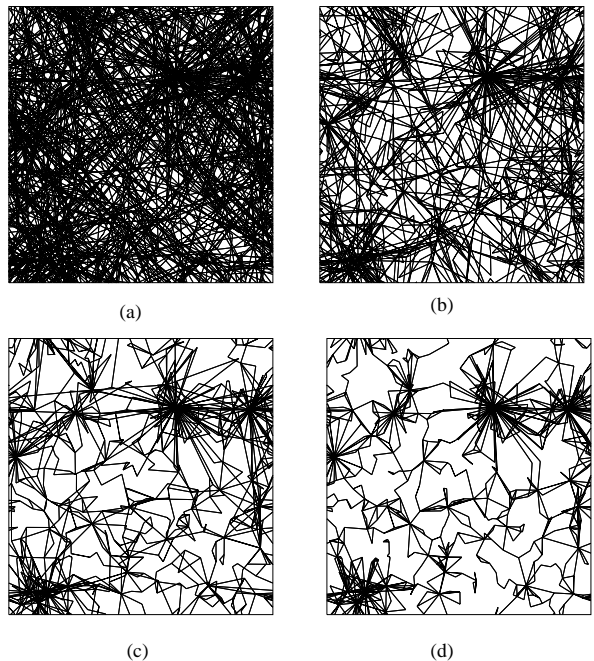}
\end{center}
\caption{
  Snapshots of the optimized network generated by the random rewiring process for
  $N$=512 and $m$=2. (a) The initial 
  network with ${\cal L} \approx 395$ (b) ${\cal L} \approx 304$ at $t$=1000 (c) 
  $ {\cal L} \approx 170$ at $t$=10000 and (d) $ {\cal L} \approx 68$ at $t$=10000000
  where $t$ is the number of rewiring trials.
}
\label{Fig2}
\end{figure}

    Let ${\bf a}$ denote the symmetric adjacency matrix of size $N \times N$ for our 
  network such that $a_{ij}=1$ if there is a link between the pair of nodes $i$ and $j$
  and 0 otherwise. Let $\ell_{ij}$ denote the shortest Euclidean distance between the 
  pair of nodes $i$ and $j$ taking into account the periodic boundary condition. Therefore when
  $a_{ij}=1$, $\ell_{ij}$ is the length of the connecting wire of the link between $i$ and $j$.
  The total cost function ${\cal L}(N)$ is therefore the sum over all link lengths
  of the network i.e., ${\cal L}(N) = \Sigma_{i>j} a_{ij}\ell_{ij}$. 

    For the convenience of discussion we define a degree vector similar to the contact vector
  generally used in the polymer physics. Our degree vector ${\bf c}$ describes the topological 
  connectivity of the network and has $N$ elements $c_i = k_i$, the degree of the $i$-th node.
  In the initial BA scale-free network one can associate a notion of time as if
  nodes are introduced one at each time unit. Therefore the links of the node $i$
  introduced at time $i$ are divided into two groups `outgoing' and `incoming'.
  Each node has only $k^{out}=m$ outgoing links connected to $m$ other nodes which are
  older than this node and it can be connected to $k^{in}=k-m$ other nodes which are
  younger than this node. Consequently the degree vector can be split into two
  other degree vectors ${\bf c^{out}}$ and ${\bf c^{in}}$ such that $c_i^{out}=k_i^{out}$
  and $c_i^{in}=k_i^{in}$ and ${\bf c=c^{out}+c^{in}}$.

\begin{figure}[top]
\begin{center}
\includegraphics[width=7.5cm]{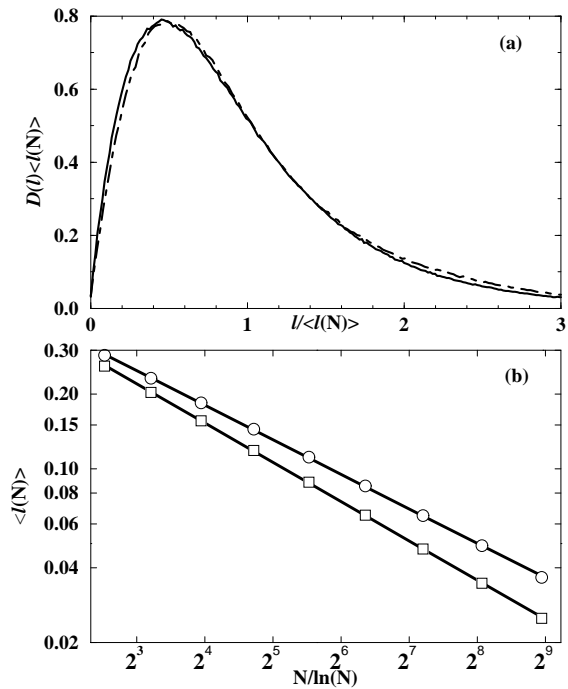}
\end{center}
\caption{
  (a) The link length distribution $D(\ell)$ in the optimized network: time-ordered (solid 
  line) and random (dashed line) scaled by the average link length $\langle \ell (N) \rangle$
  for $N$=1024.
  (b) The average length $\langle \ell (N) \rangle$ varies as $(N/\log N)^{\mu}$ 
  in the optimized network: time-ordered (circles) and random (squares) rewirings.
}
\label{Fig3}
\end{figure}

    Next, we perform the optimization dynamics to minimize the total cost function ${\cal L}(N)$.
  The optimization dynamics conserves the number of links in the network, in addition it not only 
  maintains the same degree vector ${\bf c}$ but also ${\bf c^{out}}$
  and ${\bf c^{in}}$ separately and thus ensures that the degree
  distribution of the network remains exactly same as it is before the optimization process starts.
  We call it as `time-ordered' rewiring.
  One trial of rewiring in the optimization scheme consists of selecting four nodes $n_1, n_2, n_3$ and $n_4$.
  The first node $n_1$ is randomly selected from the set of $N$ nodes. $n_2$ is selected randomly
  from the $k_1$ neighbours of $n_1$. Similarly $n_3 (\ne n_1 \ne n_2)$ is selected randomly
  from $N$ nodes and $n_4 (\ne n_1 \ne n_2)$ is again one of the $k_3$ neighbours of $n_3$.
  The move must maintain the conservation of link numbers as well as degree distribution.
  We replace the link pair $n_1n_2$ and $n_3n_4$ by another pair of links if either of the following two conditions
  is satisfied:
\begin{itemize}
\item [i.]  if $a_{13} = a_{24} =0$ and $\ell_{12}+\ell_{34} > \ell_{13}+\ell_{24}$ we link $n_1n_3$ and $n_2n_4$.
\item [ii.] if $a_{14} = a_{23} =0$ and $\ell_{12}+\ell_{34} > \ell_{14}+\ell_{23}$ we link $n_1n_4$ and $n_2n_3$.
\end {itemize}
  If both are satisfied we accept one of them with probability 1/2. If only one is
  satisfied we accept that (Fig. 1). If none of the two is satisfied we go for a fresh trial. 
  We also study a second type of rewiring process where only the total degree vector ${\bf c}$ 
  is maintained but not individually ${\bf c^{out}}$ and ${\bf c^{in}}$. Here in the final
  optimized network a particular node may have all neighbours which are younger than this node.
  We call this process as the `random' rewiring method.
  In Fig. 2 we show how an initially complicated network becomes less messy with
  increasing number of rewiring trials.
  
\begin{figure}[top]
\begin{center}
\includegraphics[width=8.5cm]{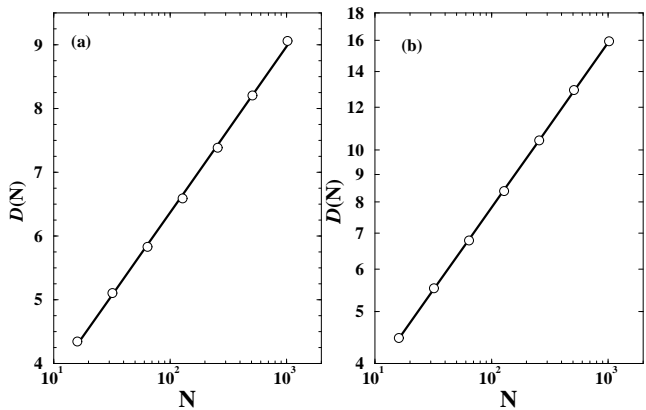}
\end{center}
\caption{
The average diameter of the network ${\cal D}(N)$ after optimization, as 
a function of number of nodes $N$ in the network:
(a) for the time-ordered exchange ${\cal D}(N) = A + B \log N$
where $A$=1.22 and $B$=1.11 where as
(b) For the random exchange ${\cal D}(N) \sim N^{\nu}$ where $\nu = 0.31 \pm 0.04$.
}
\label{Fig4}
\end{figure}

    Since we accept the move only if the total rewired cost is reduced the trial is similar to
  the zero temperature Monte Carlo dynamics. The total cost ${\cal L}(N)$ monotonically decreases
  with the number of successful trials and the number of un-successful trials between successive
  accepted moves increases. We typically try around $10(mN)^2$ such trials
  so that the plot of ${\cal L}(N)$ with logarithm of the number of trials nearly reaches a plateau.

    From each point one can measure $N-1$ distances and if these
  distances are sorted in an increasing sequence, one has the
  first neighbour distance, second neighbour distance, ...
  $(N-1)$-the neighbour distance etc. It is known that the average
  $n$-th neighbour distance $\langle R^n_N \rangle$ varies as $N^{-1/2}$ if $n/N<<1$
  and it is of the order of 1 when $n/N \sim 1$ in the
  limit of $N \to \infty$ \cite {BKC}. There is no other variation like $N^{-x}$ when
  $x$ is neither 0 nor 1/2 but in between.

  In the optimized network the links are not necessarily a fixed $(n)$ neighbour distances
  but a complex mixture of many neighbour distances. More 
  elaborately it is expected that many of the links of the optimized
  network are first neighbour distances, less numbers are second neighbour 
  distances, less numbers are third neighbour distances etc. 
  In the optimized network we first calculate the probability density of the link length distribution
  $D(\ell)$. This distribution on scaling by the average link length $\langle \ell (N) \rangle$
  is nearly the same for the time-ordered as well as the random rewiring processes.
  Contrary to the expectation this distribution has a maximum and it fits very 
  well to a functional form $D(\ell)\langle \ell (N) \rangle \sim x^{\alpha}e^{-ax^{\beta}}$
  with $x=\ell/\langle \ell (N) \rangle$. The fit on a linear scale gives
  $\alpha = 1.4, 1.1$ and $\beta = 0.8, 0.9$ approximately
  for the time-ordered and random rewiring processes respectively.
  The network has $N_{\ell} = 2N-m-1$ links and therefore ${\cal L}(N)=N_{\ell} \langle \ell (N) \rangle$.
  We plot in Fig. 3
  $\langle \ell (N) \rangle$ with $N/\log N$ and observe excellent straight lines on a
  double logarithmic scale. Therefore $\langle \ell (N) \rangle \sim (N/\log N)^{\mu}$ 
  where $\mu = 0.46$ and 0.52 with an error of 0.05 approximately 
  for the time-ordered and random rewiring processes respectively.

\begin{figure}[top]
\begin{center}
\includegraphics[width=6.5cm]{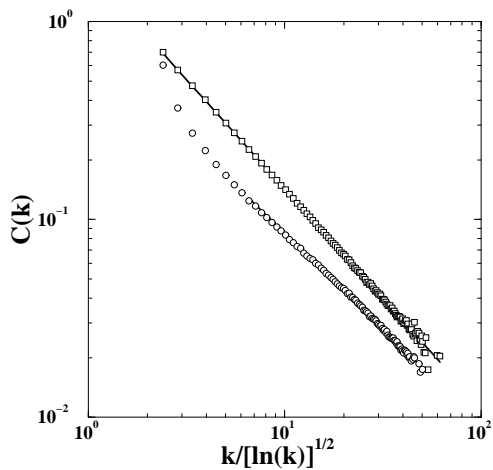}
\end{center}
\caption{
The clustering co-efficient ${\cal C}(k)$ as a function of the degree $k$
shows a $\{k/\{ln(k)\}^{1/2}\}^{-b}$ behaviour with $b \approx 0.94$ and 1.1
for the time-ordered (circles) and random (squares) rewiring processes respectively.
}
\label{Fig5}
\end{figure}

    The topological size of the network is measured by the diameter of the network.
  The distance ${\it d}_{ij}$ between an arbitrary pair of nodes $i$ and $j$ is the number of links on the shortest
  path connecting the two nodes. The diameter $d_m$ is the maximal distance on a network.
  The average diameter ${\cal D}(N)$ represents the configuration averaged maximal distance $\langle d_m \rangle$.
  Variations of the average diameter of the optimized network with the network size
  is shown in Fig. 4. For the the time-ordered exchange ${\cal D}(N) = A + B\log N$ 
  with $A \approx 1.22$ and $B \approx 1.11$ where as for the random exchange 
  ${\cal D}(N) \sim N^{\nu}$ with $\nu = 0.31 \pm 0.04$
  where the error 0.04 is estimated by the largest difference of the local slopes between
  successive points and the mean slope.
  Although the degree distribution remains scale-free in the optimized networks generated by both the time-ordered
  and in random rewiring procedures, the first network retains some long
  range connections due to the constraint that both ${\bf c^{out}}$ and ${\bf c^{in}}$
  are strictly maintained where as in the second network by random rewirings
  essentially all the connections are local, i.e. typically a node has all neighbours
  within a spatial distance of the order of $\sim N^{-1/2}$. 

    The local correlation among the links is measured by the clustering co-efficient.
  The clustering co-efficient ${\cal C}_i$ of the $i$-th node is measured by the
  ratio of the number of links $e_i$ within the $k_i$ neighbours of the $i$-th node
  and the number of links $k_i(k_i-1)/2$ if the $k_i$ nodes have formed an $k_i$-clique
  i.e., ${\cal C}_i = 2e_i/\{k_i(k_i-1)\}$. The clustering co-efficient of the whole
  network ${\cal C}(N)$ is $\langle {\bar C} \rangle$. Also the average clustering co-efficient
  for the set of nodes of degree $k$ is defined as ${\cal C}(k)$. In general both these 
  clustering co-efficients may decrease as power laws: ${\cal C}(N) \sim N^{-a}$ and
  ${\cal C}(k) \sim k^{-b}$. In our case we start from the initial BA network where
  it is known that $a \approx 3/4$ and $b = 0$ \cite {BA}. We also calculate these quantities
  in the final optimized state. The total clustering coefficient is found to be independent
  of $N$ and therefore $a$=0 where as unlike a simple power law for $C(k)$ we get a power law with
  logarithmic correction. In Fig. 5 we plot ${\cal C}(k)$ with $k/\{ln(k)\}^{1/2}$
  and observe straight lines on a double logarithmic scale implying the variation as:
\begin {equation}
{\cal C}(k) \sim \{k/\{ln(k)\}^{1/2}\}^{-b}
\end {equation}
  where  $b \approx 0.94$ and 1.1 for the time-ordered and random rewiring processes respectively.
  We cannot rule out the possibility that $b$ is actually 1 for both the processes.
  Many networks and models show $b=1$ \cite {Janos,BA}.

    To summarize, we have studied a  cost optimized network which has three main features of
  the real-world networks e.g., it is a small-world network, it is a scale-free network
  and also it exhibits high clustering properties as well.
  We studied this network on the two-dimensional Euclidean space which should be relevant
  in the context of the Internet. While some links in Internet are the
  cable-less (microwave) links, many connections are made by real physical Ethernet cables.
  Therefore the question of optimizing cost of the total wiring length of the
  network arises naturally which is the main point of study in this paper. An
  optimized geographical embedding algorithm for scale-free networks was recently
  studied independently \cite {Rozenfeld}. Unlike in \cite {Rozenfeld}, our time-ordered optimization
  produces a statistically non-homogeneous network and preserves a significant
  number of long-distance connections, permitting the network diameter to still
  scale as $\log(N)$ as $N \to \infty$. We also obtain a stretched-exponentially decaying
  tail of the link length distribution in the optimized network which is unlike the 
  power-law tail observed by Yook et.al. \cite {Yook} and closer to the Waxman result \cite {Waxman}.

    SSM thankfully acknowledges P. Sen, P. Bhattacharyya and J. Kert\'esz for critical reading of the manuscript
  and the Dipartimento di Fisica, University of Padova for hospitality.
  AK gratefully acknowledges the financial support of MIUR through COFIN 2001.


\begin{thebibliography}{90}
\bibitem {Faloutsos} M. Faloutsos, P. Faloutsos and C. Faloutsos, Proc.
		ACM SIGCOMM, Comput. Commun. Rev., {\bf 29}, 251 (1999).

\bibitem {web} S. Lawrence and C. L. Giles, Science, {\bf 280}, 98 (1998); Nature, {\bf 400}, 107 (1999),
               R. Albert, H. Jeong and A.-L. Barab\'asi, Nature, {\bf 401}, 130 (1999).

\bibitem {Newman} M. E. J. Newman, {\it Proc. Natl. Aca. Sci.}, {\bf 98}, 404 (2001);arXiv:cond-mat/0011155.

\bibitem {Jeong} H. Jeong, S. P. Mason,  A.-L. Barab\'asi and Z. N. Oltvai, Nature, {\bf 411}, 41 (2001).

\bibitem {Sole} R. V. Sol\'e and J. M. Montoya, Proc. Royal Soc. London B, {\bf 268}, 2039 (2001);
J. Camacho, R. Guimer\'a and L. N. Amaral, Phys. Rev. Lett. {\bf 88}, 228102 (2002).

\bibitem {WS} D. J. Watts and S. H. Strogatz, Nature, {\bf 393}, 440 {1998};
D. J. Watts, {\it Small Worlds: The Dynamics of Networks Between order and Randomness},
(Princeton 1999).

\bibitem {BA} A.-L. Barab\'asi and R. Albert, Science, {\bf 286}, 509 (1999);
R. Albert and A.-L. Barab\'asi, Rev. Mod. Phys. {\bf 74}, 47 (2002).

\bibitem {Waxman} B. Waxman, IEEE J. Selec. Areas Commun., SAC, {\bf 6}, 1617 (1988).

\bibitem {Yook} S.-H. Yook, H. Jeong and A.-L. Barab\'asi, arXiv:cond-mat/0107417.

\bibitem{MS} S. S. Manna and P. Sen, Phys. Rev. E {\bf 66}, 066114 (2002).

\bibitem{MS1} P. Sen and S. S. Manna, 2003, arXiv:cond-mat/0301617.

\bibitem{Xulvi} R. Xulvi-Brunet and I. M. Sokolov, arXiv:cond-mat/0205136.

\bibitem{Barthelemy} M. Barth\'elemy, arXiv:cond-mat/0212086.

\bibitem{Rozenfeld} A. F. Rozenfeld, R. Cohen, D. b-Avraham and S. Havlin, Phys. Rev. Lett.
{\bf 89}, 218701 (2002); D. b-Avraham, A. F. Rozenfeld, R. Cohen and S. Havlin, arXiv:cond-mat/0301504.

\bibitem {BKC} P. Bhattacharyya, B. K. Chakrabarti, math.PR/0212230.

\bibitem {Janos} G. Szab\'o, M. Alava and J. Kert\'esz, arXiv:cond-mat/0208551.
\end{thebibliography}
\end{document}